\documentclass[journal]{IEEEtran}
\usepackage{amsmath,graphicx,cite,epsfig}
\usepackage{multirow}
\usepackage{booktabs}
\usepackage{float}
\usepackage{subfigure}
\usepackage{caption}
\usepackage{threeparttable}
\usepackage{diagbox}
\usepackage{url}
\usepackage{bm}
\usepackage{stfloats}
\usepackage{amsfonts}
\usepackage{color}

\ifCLASSINFOpdf

\else

\fi

\begin{document}
%
% paper title
% Titles are generally capitalized except for words such as a, an, and, as,
% at, but, by, for, in, nor, of, on, or, the, to and up, which are usually
% not capitalized unless they are the first or last word of the title.
% Linebreaks \\ can be used within to get better formatting as desired.
% Do not put math or special symbols in the title.
\title{Blind Quality Assessment of 3D Dense Point Clouds with Structure Guided Resampling}
%
%
% author names and IEEE memberships
% note positions of commas and nonbreaking spaces ( ~ ) LaTeX will not break
% a structure at a ~ so this keeps an author's name from being broken across
% two lines.
% use \thanks{} to gain access to the first footnote area
% a separate \thanks must be used for each paragraph as LaTeX2e's \thanks
% was not built to handle multiple paragraphs
%

\author{Wei Zhou, Qi Yang, Qiuping Jiang, Guangtao Zhai,~\IEEEmembership{Senior Member, IEEE}, and Weisi Lin,~\IEEEmembership{Fellow, IEEE}% <-this % stops a space
\thanks{This work was supported in part by NSFC under Grant 61901236 and the Natural Science Foundation of Zhejiang LR22F020002. (Corresponding author: Qiuping Jiang.)}
\thanks{W. Zhou is with the Department of Electrical and Computer Engineering, University of Waterloo, Waterloo, ON N2L 3G1, Canada (e-mail: wei.zhou@uwaterloo.ca).}
% <-this % stops a space
% <-this % stops a space
\thanks{Q. Yang is with the Tencent MediaLab, Shanghai 200030, China (e-mail:
chinoyang@tencent.com).}
\thanks{Q. Jiang is with the School of Information Science and Engineering, Ningbo University, Ningbo 315211, China (e-mail: jiangqiuping@nbu.edu.cn).}
\thanks{G. Zhai is with the Institute of Image Communication and Network Engineering, Shanghai Jiao Tong University, Shanghai 200240,
China (e-mail: zhaiguangtao@sjtu.edu.cn)}
\thanks{W. Lin is with the School of Computer Science and Engineering,
Nanyang Technological University, Singapore 639798 (e-mail:
wslin@ntu.edu.sg).}}

\maketitle

% As a general rule, do not put math, special symbols or citations
% in the abstract or keywords.
\begin{abstract}
Objective quality assessment of 3D point clouds is essential for the development of immersive multimedia systems in real-world applications. Despite the success of perceptual quality evaluation for 2D images and videos, blind/no-reference metrics are still scarce for 3D point clouds with large-scale irregularly distributed 3D points. Therefore, in this paper, we propose an objective point cloud quality index with Structure Guided Resampling (SGR) to automatically evaluate the perceptually visual quality of 3D dense point clouds. The proposed SGR is a general-purpose blind quality assessment method without the assistance of any reference information. Specifically, considering that the human visual system (HVS) is highly sensitive to structure information, we first exploit the unique normal vectors of point clouds to execute regional pre-processing which consists of keypoint resampling and local region construction. Then, we extract three groups of quality-related features, including: 1) geometry density features; 2) color naturalness features; 3) angular consistency features. Both the cognitive peculiarities of the human brain and naturalness regularity are involved in the designed quality-aware features that can capture the most vital aspects of distorted 3D point clouds. Extensive experiments on several publicly available subjective point cloud quality databases validate that our proposed SGR can compete with state-of-the-art full-reference, reduced-reference, and no-reference quality assessment algorithms.
\end{abstract}

% Note that keywords are not normally used for peerreview papers.

% For peer review papers, you can put extra information on the cover
% page as needed:
% \ifCLASSOPTIONpeerreview
% \begin{center} \bfseries EDICS Category: 3-BBND \end{center}
% \fi
%
% For peerreview papers, this IEEEtran command inserts a page break and
% creates the second title. It will be ignored for other modes.
\IEEEpeerreviewmaketitle

\section{Introduction}
% The very first letter is a 2 line initial drop letter followed
% by the rest of the first word in caps.
%
% form to use if the first word consists of a single letter:
% \IEEEPARstart{A}{demo} file is ....
%
% form to use if you need the single drop letter followed by
% normal text (unknown if ever used by the IEEE):
% \IEEEPARstart{A}{}demo file is ....
%
% Some journals put the first two words in caps:
% \IEEEPARstart{T}{his demo} file is ....
%
% Here we have the typical use of a "T" for an initial drop letter
% and "HIS" in caps to complete the first word.

\IEEEPARstart{W}{ith} the modern advances in 3D data capture devices and rendering technologies, popular 3D point clouds have become one of the most important multimedia representations for providing 6 degrees of freedom \cite{van2019towards}. As defined, a 3D point cloud contains a huge number of scattered points with certain attributes. Each 3D point owns geometry and attribute information, which can represent actual objects or environments more visually compared with traditional images and videos. To be specific, geometry stands for the 3D coordinates of the point, while the attribute is composed of additional descriptors such as RGB color, surface normal, reflectance and opacity, etc. Due to such an abundant pattern, there have emerged lots of real-world applications for 3D point clouds, including VR/AR/XR \cite{wang2018point}, automatic diving \cite{yue2018lidar}, scene understanding \cite{chen2019deep}, among many others. Typically, 3D point clouds can be divided as two types: relatively smaller objects and large-scale scenes. In this work, we focus on the first type of 3D dense point clouds.

\begin{figure}[t]
	\centerline{\includegraphics[width=9.2cm]{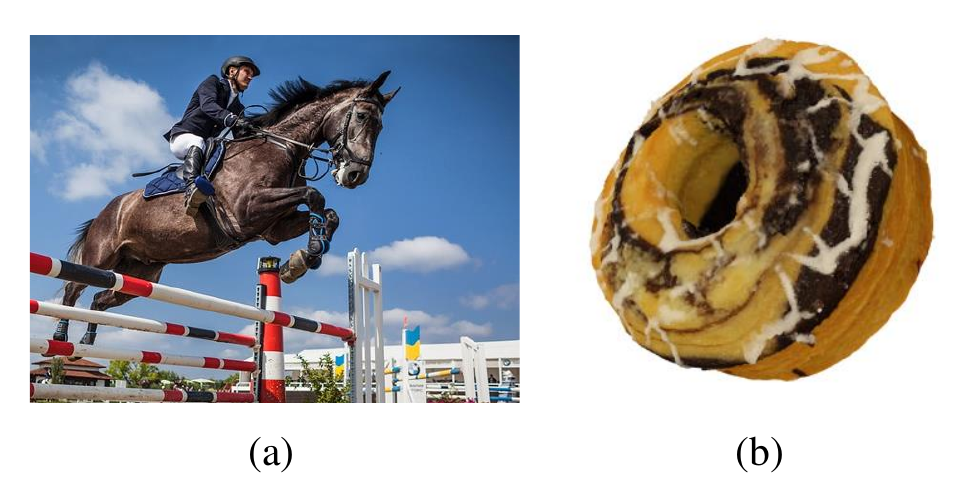}}
	\caption{Comparison between a 2D image and a 3D point cloud. (a) 2D image, (b) 3D point cloud.}
	\centering
	\label{figure1}
\end{figure}

Although the data formats of 3D point clouds are very different, similar to conventional images and videos, a variety of distortions would be inevitably introduced during the processing chain of multimedia communication systems \cite{sun2022graphiqa}. For example, the acquisition sensors may produce noise artifacts in captured 3D point clouds. Moreover, because of the large-scale volumetric content of 3D point clouds, executive downsampling and compression operations \cite{schwarz2018emerging} can also lead to perceptual quality degradation. Consequently, how to effectively assess the perceptually visual quality of 3D point clouds is a significant and challenging problem, which can benefit other relevant processing stages, e.g. tuning the acquisition parameters for the optimal capture of 3D point clouds. Regarding to the quality assessment of 3D point clouds, a subjective test usually serves as the most accurate and reliable method \cite{zhang2014subjective,alexiou2017towards}. However, since organizing subjective experiments are time-consuming and labor-intensive, developing efficient objective quality metrics is a promising alternative.

In general, the characteristics of point clouds make the design of objective quality assessment models more challenging than traditional 2D images or videos. As shown in Fig. \ref{figure1}, here we illustrate two examples of a 2D image and a 3D point cloud, where the former is a man riding on a horse and the latter is a cake as an object. Ideally, both the two data formats can represent a scene or an object in real-world. However, we can see that a 2D image is usually displayed in a regular shape, while a 3D point cloud is unstructured and composed of many distributed points. Thus, new challenges would arise for the objective quality assessment of 3D point clouds.

\begin{figure*}[t]
	\centerline{\includegraphics[width=17.8cm]{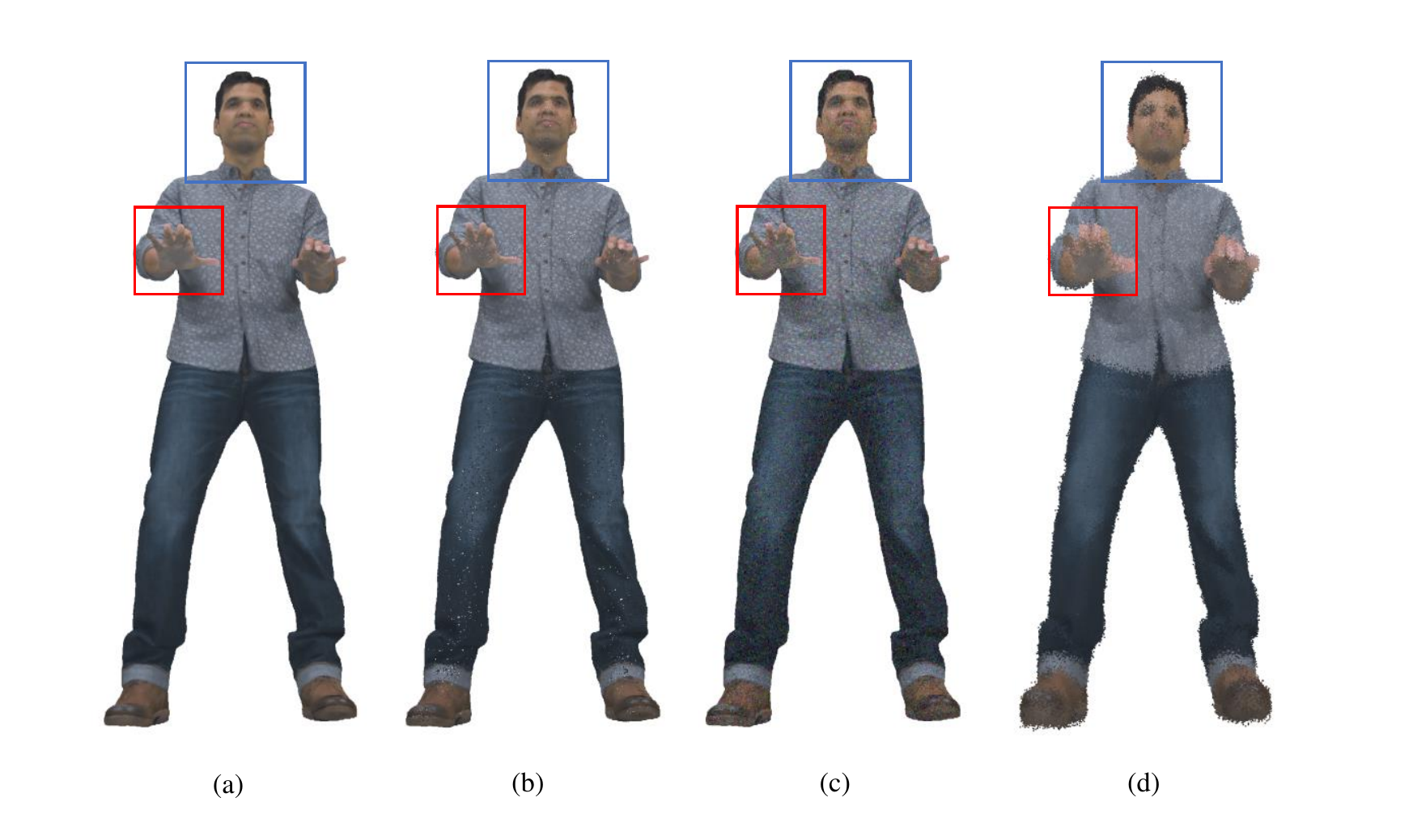}}
	\caption{Typical distortions in 3D point clouds. (a) A reference point cloud, (b-d) examples of distorted point clouds.}
	\centering
	\label{distortion}
\end{figure*}

According to the availability of original reference 3D point clouds, the objective quality assessment methods of 3D point clouds can be divided into three categories, consisting of full-reference (FR), reduced-reference (RR), and no-reference (NR) models. That is, the FR and RR models can access full and part data of reference point clouds, respectively. In contrast, the blind/NR models evaluate the perceptual quality of distorted 3D point clouds without any reference information. Among the NR models, general-purpose ones can evaluate distortion-generic 3D point clouds, which do not need to know about the distortion type and are more practical in real-world scenarios. Existing general-purpose NR quality assessment methods for 3D point clouds have been proposed mainly on the basis of deep neural networks (DNNs) \cite{liu2021pqa,yang2022no,liu2020point}. By leveraging the remarkable learning ability of DNNs, perceptual quality can be directly predicted from the point data. However, the learning process often lacks the design philosophies of domain knowledge that reflect the human visual system (HVS) characteristics, causing unsatisfactory interpretability. In \cite{zhang2022no}, a statistics-based quality assessment model for point clouds and meshes was proposed. To the best of our knowledge, so far there has been no exploration of explainable general-purpose NR quality assessment methods specifically designed for 3D point clouds based on domain knowledge.

As illustrated in Fig. \ref{distortion}, our main motivation comes from the observations of typical artifact appearances in various distorted 3D point clouds. From (b) to (d), we can find the perceptual quality degradations are relevant to geometry density, color difference and point orientation, respectively. For example, compared with the original reference point cloud in figure (a), the points are incomplete and have incorrect color in figures (b) and (c), respectively. Moreover, in figure (d), the orientation of these points are destroyed, as clearly observed by the object parts in the bounding boxes. Therefore, we aim to expose the black box of point cloud quality prediction by quantifying the distortions from the aspects of density, color and orientation, leading to the general-purpose NR point cloud quality index with Structure Guided Resampling (SGR).

In our proposed SGR, to extract efficient HVS-related features based on scattered points, regional pre-processing is first conducted, where keypoint resampling and local region construction are involved. In this pre-processing step, considering the huge amount of point clouds data and the crucial structure information, we first resample the test point cloud via normal vectors to obtain a serial of keypoints, and then construct local regions centered keypoints. Afterwards, inspired by the combined impacts of 3D geometry and associated attributes, 3 groups of quality-aware features including 1) geometry density features, 2) color naturalness features, and 3) angular consistency features, are employed to predict the perceptual quality of distorted 3D point clouds. Besides, we also consider the fundamental theory of natural scene statistics (NSS) \cite{ruderman1994statistics,srivastava2003advances} in the extrated features. Finally, the quality-related features are fused into the overall SGR via a quality regression module. The effectiveness of our proposed SGR is verified on four subjective point cloud quality databases. The experiments demonstrate that the proposed SGR presents significant performance improvement compared with other NR metrics, and even better than most FR metrics. Besides, the proposed SGR shows stable performance with typical point cloud distortion types, such as Downsampling, Gaussian Noise, G-PCC, and V-PCC.

The main contributions of this paper are summarized as follows:

\begin{itemize}
\item We propose a general-purpose blind objective quality assessment algorithm for 3D point clouds based on the cognitive characteristics of the human brain and the regularity of NSS.
\item According to the structural dependence of the HVS, the regional pre-processing is proposed in the SGR framework, which exploits the unique normal vectors of 3D point clouds.
\item The integrated influence of 3D geometry and associated attributes information can be reflected by the proposed quality-aware features. Experimental results on existing subjective point cloud quality databases demonstrate the competitive performance of the proposed model.
\end{itemize}

\begin{figure*}[t]
	\centerline{\includegraphics[width=18.5cm]{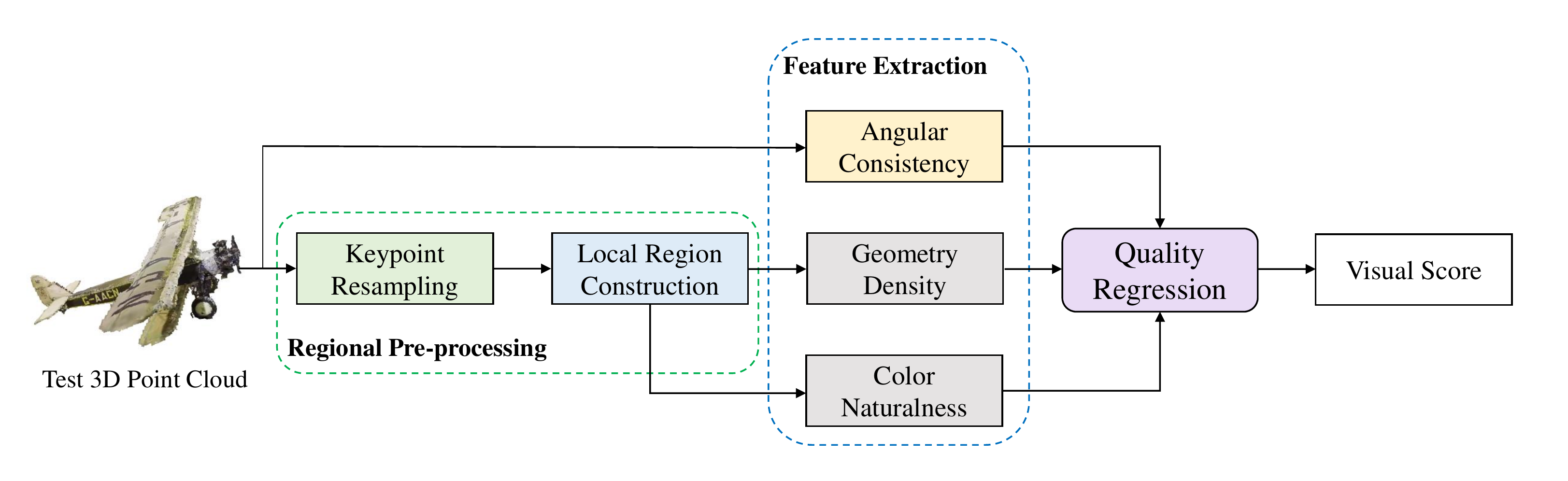}}
	\caption{Framework of the proposed SGR method which consists of regional pre-processing, quality-aware feature extraction, and quality regression. The regional pre-processing involves keypoint resampling and local region construction. For final visual score prediction, geometry density, color naturalness and angular consistency features are used to quantify the perceptual quality of 3D point clouds.}
	\centering
	\label{figure2}
\end{figure*}

The rest of this paper is organized as follows. Section II reviews related work on objective quality assessment of 3D point clouds. In Section III, we introduce the technical details of our proposed NR quality assessment method. Section IV presents the validation results of the proposed model. Finally, we conclude the paper in Section V.

\section{Related Work}
To assess the perceptual quality of distorted 3D point clouds, comparing with the corresponding original reference point clouds is the most intuitive way if the pristine one is available. Therefore, there have emerged many FR quality assessment methods for 3D point clouds, which can be generally classified into point-based and projection-based metrics.

The point-based models directly use the 3D points in the reference and distorted contents. Among them, the earliest point-based metric is point-to-point (p2po) \cite{mekuria2016evaluation}, where representative mean squared error (MSE) or Hausdorff (HF) distance is often adopted for geometric peak signal-to-noise ratio (PSNR) computation. An alternative method is point-to-plane (p2pl) \cite{tian2017geometric} that computes the PSNR between a test point and a corresponding normal vector in the reference point cloud. Except for the point-based geometry distortion measures, color information can also be used to compute the PSNR, such as the YUV channels or the luminance Y component \cite{mekuria2017performance}. But both p2po and p2pl quality metrics cannot precisely evaluate the perceptual quality of 3D point clouds under structural loss. Thus, the angular similarity (AS), also known as plane-to-plane, was proposed \cite{alexiou2018point}. Moreover, on the basis of graph signal processing, Yang et al. \cite{yang2020inferring} proposed graph similarity index (GraphSIM) and the extended multiscale GraphSIM (MS-GraphSIM) \cite{zhang2021ms}. In addition, based on the designed mesh structural distortion measure (MSDM) \cite{lavoue2006perceptually}, Meynet et al. \cite{meynet2019pc} exploited local curvature statistics to develop the FR quality assessment method called PC-MSDM for the perceptual quality evaluation of 3D point clouds. They further proposed the point cloud quality metric (PCQM) \cite{meynet2020pcqm} which utilizes the optimally weighted linear combination of geometric curvature and color features. Furthermore, similar to the well-known structural similarity index (SSIM) \cite{wang2004image}, Alexiou et al. \cite{alexiou2020towards} tried to capture the local changes of test point clouds, leading to the quality degradation measurement named PointSSIM.

Apart from the previously mentioned point-based FR methods, another kind of mainstream FR framework is to project 3D point clouds into multiple 2D images from various views and then conduct the 2D image quality assessment, namely projection-based models. For example, the SSIM \cite{wang2004image} and its variant multiscale SSIM (MS-SSIM) \cite{wang2003multiscale} as well as the visual information fidelity in the pixel domain (VIFP) \cite{sheikh2006image} are often employed to predict the visual quality of converted 2D images. Several projection methods can be adopted, where 6 perpendicular projection of a cube is one of typical ways \cite{yang2020predicting}. Usually, different weights are allocated for extracted features from all projection planes.

Since the full information of associated reference point clouds do not always exist, Viola et al. \cite{viola2020reduced} proposed a RR quality metric for point cloud contents (i.e. PCMRR), where a small set of features from the references are extracted and then delivered to the receiver side for evaluating the quality degradation of distorted point clouds. Additionally, some works also use DNNs to explore the challenging NR quality metrics of 3D point clouds. For example, Liu et al. \cite{liu2021pqa} designed a point cloud quality assessment network (PQA-Net) which consists of multi-view-based joint feature extraction and fusion, distortion type identification, and final quality prediction. More recently, adversarial domain adaption has been used for the development of NR quality assessment methods for point clouds \cite{yang2022no}.

However, the existing NR models are generally data-driven and uninterpretable, little consideration is given to design domain knowledge oriented perceptual features that reveal the characteristics of the HVS. Therefore, in this work, we aim to bridge the explicit gap between the HVS perception and NR point cloud quality assessment. Specifically, considering the structure information of scattered points and the great success of NSS regularity in 2D image quality prediction, we are the first to propose a blind/NR quality assessment method specifically designed for 3D point clouds based on well-defined quality-aware features upon regional pre-processing and the cognitive properties of the human brain.

\begin{figure*}[t]
	\centerline{\includegraphics[width=18.2cm]{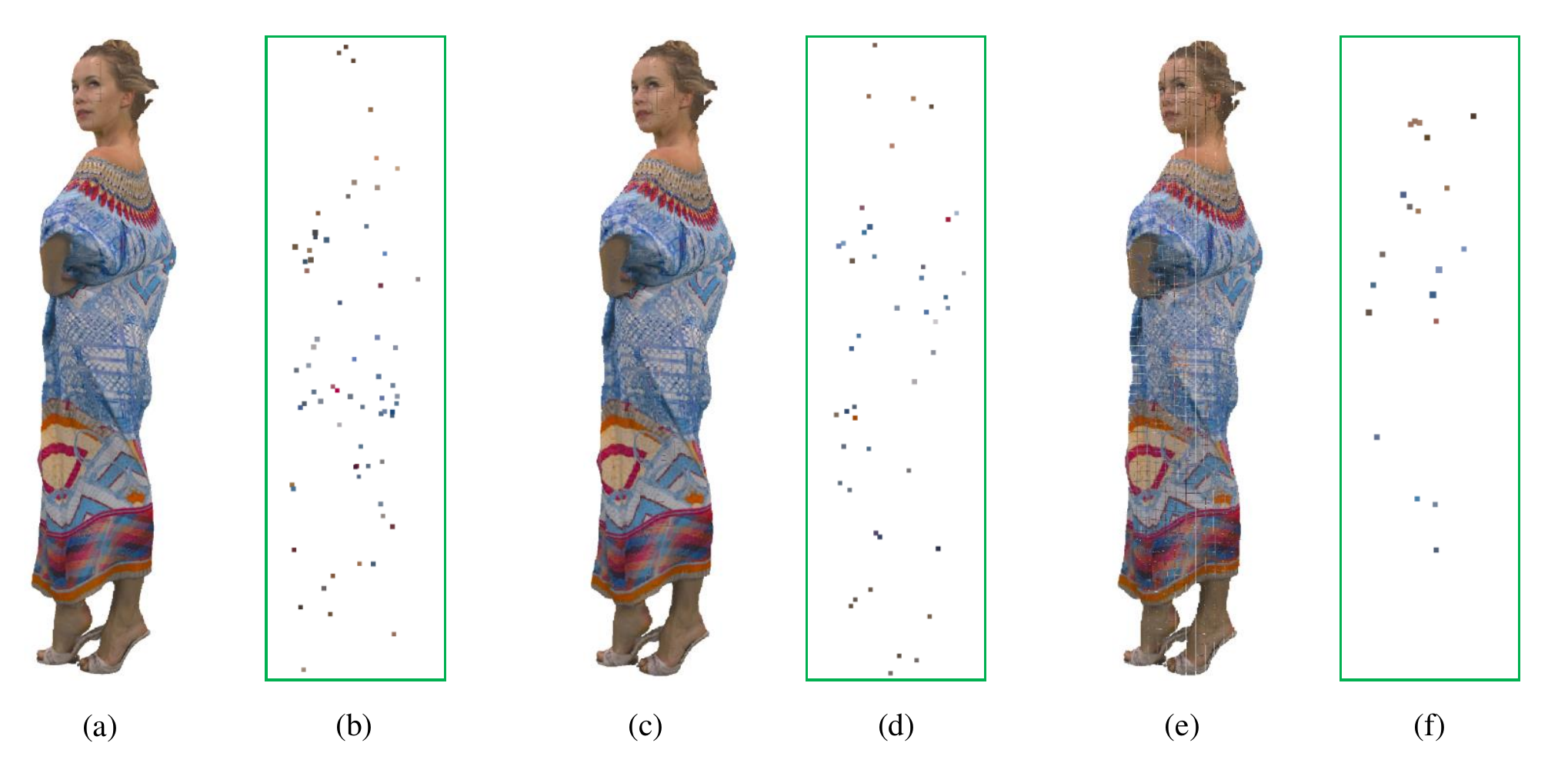}}
	\caption{Illustration of keypoint resampling for 3D point clouds under various distortion levels. (a)(c)(e) Distorted 3D point clouds, (b)(d)(f) the corresponding keypoints resampled from (a)(c)(e).}
	\centering
	\label{figure3}
\end{figure*}

\section{Proposed Point Cloud Quality Index}
The framework of our proposed SGR method is shown in Fig. \ref{figure2}, which is a general-purpose blind quality assessment model.  Since the HVS is more sensitive to structures, we first resample keypoints from the input test 3D point cloud based on generated normal vectors. Then, we construct local regions for the resampled points and extract distortion-related features. Finally, the quality regression module is used to map the extracted features onto visual scores. We will introduce the technical details of SGR in the following subsections.

\subsection{Regional Pre-processing}
\textbf{Keypoint resampling.} Suppose that a distorted 3D point cloud $\textbf{P}$ contains $\rm{M}$ points and each point has three geometry coordinates (e.g., $x, y, z$) as well as three color attributes (e.g., $r, g, b$), we can denote the 3D point cloud data as follows:

\begin{equation}
\textbf{P}=\left[\textbf{p}_{1}, \textbf{p}_{2}, \ldots, \textbf{p}_{\rm{M}}\right]^{T} \in \mathbb{R}^{\rm{M} \times 6},
\end{equation}
where $\textbf{p}_{m}=\left(x_{m}, y_{m}, z_{m}, r_{m}, g_{m}, b_{m}\right)$ is a point in the point cloud. In this representation, the geometry coordinates and color attributes can be separated as $\textbf{g}_{m}=\left(x_{m}, y_{m}, z_{m}\right)$ and $\textbf{c}_{m}=\left(r_{m}, g_{m}, b_{m}\right)$, respectively. In addition, another attribute called normal vectors can be estimated from the point cloud.

Generally, each 3D point cloud data has numerous points. For example, the 3D point cloud in Fig. \ref{figure1} (b) has $\rm{M}=2,486,566$ points. Moreover, these 3D points are irregular and unstructured. Nevertheless, the HVS inclines to perceive the structures of objects or environments. For traditional images and videos, such structure information has been widely used to design many objective quality measures \cite{wang2004image,wang2003multiscale,wang2010information}. Inspired by GraphSIM \cite{yang2020inferring}, we opt to extract the keypoints by graph-based resampling to realize the structure extraction of 3D point clouds. Specifically, considering that the resampled points should be high-frequency parts such as edges, contours, etc, GraphSIM uses a Haar-like high pass filter to extract point cloud skeleton based on original geometry coordinates. However, this method is too sensitive to the outlier, which may lead to unstable results. Therefore, we first compute the normal vectors as and treat them as the graph signal to guide keypoint resampling.

\begin{figure}[t]
	\centerline{\includegraphics[width=5.2cm]{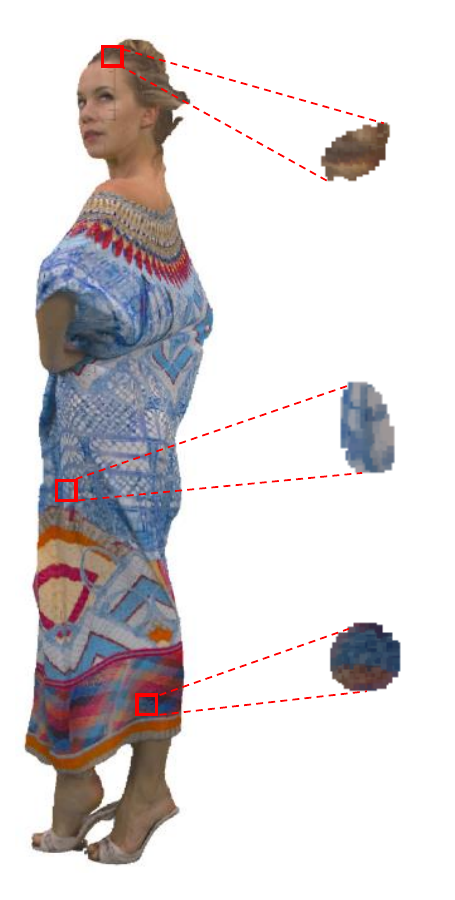}}
	\caption{Structure guided local region construction of a distorted 3D point cloud after keypoint resampling.}
	\centering
	\label{region}
\end{figure}

As defined, a graph filter represents a signal system that inputs a graph signal and outputs a tuned graph signal \cite{chen2017fast}. Mathematically, a linear and shift-invariant graph filter is a polynomial conversion of the graph shift operator $\bf{\mathcal{A}} \in \mathbb{R}^{\rm{M} \times \rm{M}}$ computed by:
\begin{equation}
h(\bf{\mathcal{A}})=\sum_{k=0}^{\rm{K}} h_{k} \bf{\mathcal{A}}^{k}=h_{0} \textbf{I}+h_{1} \bf{\mathcal{A}}+\ldots+h_{\rm{K}-1} \bf{\mathcal{A}}^{\rm{K}-1},
\end{equation}
where $h_{k}$ denotes the $\rm{k-th}$ filter coefficients and $\rm{K}$ is the length of the graph filter. $\textbf{I}$ represents the identity matrix. With the predefined graph filter, for an input graph signal $\bf{\Phi} \in \mathbb{R}^{\rm{M} \times \rm{1}}$, the filtered output graph signal is a matrix-vector product which can be converted by eigendecomposition as:

\begin{equation}
h(\bf{\mathcal{A}}) \bf{\rm{\Phi}}=\operatorname{\textbf{V}h}(\bf{\Lambda}) \textbf{V}^{-1} \bf{\Phi},
\end{equation}
where the eigenvectors of $\bf{\mathcal{A}}$ constitute the columns of matrix $\textbf{V}$. The eigenvalue matrix $\bf{\Lambda} \in \mathbb{R}^{\rm{M} \times \rm{M}}$ is the diagonal matrix containing the corresponding eigenvalues of $\bf{\mathcal{A}}$. Here, the eigenvalues are frequencies on the formed graph, which can be used to sort the 3D points.

Given a distorted 3D point cloud $\textbf{P}$, from the above analysis, we resample it to extract graph keypoints by:
\begin{equation}
\hat{\textbf{P}}=\rm{h_H}(\textbf{P}, \bf{\Phi}, \rm{K})_{\rm{\Theta}}, \rm{\Theta}<<\rm{M},
\end{equation}
where $\hat{\textbf{P}}\in \mathbb{R}^{\rm{\Theta} \times 6}$, and $\rm{h_H(\cdot)}$ is a high-pass graph filter with length equaling to $\rm{K}$, and the graph signal $\bf{\Phi}$ is normal vector. Moreover, $\rm{\Theta}$ represents the number of keypoints after the resampling operation. Similar to \cite{zhang2021ms}, we set $\rm{K}$ and $\rm{\Theta}$ as $4$ and $\rm{M}/10,000$, respectively.

As shown in Fig. \ref{figure3}, we give examples of the keypoint resampling results for 3D point clouds which are distorted by various Octree-based compression levels. From (a), (c) to (e), the distortion levels increase, which can be easily observed by the appearances of distorted point clouds. We can see that the keypoints generally represent the structures of point clouds. In addition, with the increase of distortion levels, the number of resampled points decreases. Therefore, the keypoint resampling can be a promising quality indicator for subsequent local region construction and quality-aware feature extraction. This is mainly benefit from the efficient structure information of normal attribute.

%\begin{figure}[t]
%	\centering
%	\begin{minipage}{0.49\linewidth}
%		\centerline{\includegraphics[width=4.3cm]{fig/fig3_1.pdf}}
%		\centerline{(a)}
%	\end{minipage}
%	\begin{minipage}{0.49\linewidth}
%		\centerline{\includegraphics[width=4.3cm]{fig/fig3_2.pdf}}
%		\centerline{(b)}
%	\end{minipage}
%	\caption{Illustration of normal vectors. (a) 3D point cloud, (b) Estimated normal vectors.}
%	\label{figure3}
%\end{figure}

\textbf{Local region construction.} After the process of keypoint resampling, we exploit the generated keypoints to construct local regions. For each keypoint $\hat{\textbf{p}_{\theta}}$ in $\hat{\textbf{P}}$, we cluster its neighbors based on the Euclidean distance of the corresponding geometry coordinates as follows:

\begin{equation}
\tilde{\textbf{P}} \subset \hat{\textbf{P}},\left\|\tilde{\bf{g}}-\hat{\bf{g_{\theta}}}\right\|_{2}^{2} \leq \alpha,
\end{equation}
where $\tilde{\textbf{P}}$ denotes the groups of clustered neighbors of the keypoint $\hat{\textbf{p}_{\theta}}$. $\tilde{\bf{g}}$ and $\hat{\bf{g_{\theta}}}$ are the geometry components of $\tilde{\textbf{P}}$ and $\hat{\textbf{p}_{\theta}}$, respectively. In addition, $\alpha$ is used to cluster neighbors, which is $1/20$ of minimum range among $\rm{x}$, $\rm{y}$, and $\rm{z}$ coordinates.

In Fig. \ref{region}, we show the structure guided local region construction of distorted 3D point cloud after keypoint resampling. Here, we take the figure (a) in Fig. \ref{figure3} as an example, where three constructed regions are given. It can be found that these local regions have different shapes, which are used for the quality-aware feature extraction.

% Note that the vertices of local region are the keypoints and the edge weights of local region between two vertices $\hat{\bf{p_{i}}}$ and $\hat{\bf{p_{j}}}$ are calculated by:

% \begin{equation}
% \rm{W_{i, j}}=\left\{\begin{array}{l}
% e^{-\frac{\left\|\hat{\bf{g_{i}}}-\hat{\bf{g_{j}}}\right\|_{2}^{2}}{\sigma^{2}}},\left\|\hat{\bf{g_{i}}}-\hat{\bf{g_{j}}}\right\|_{2}^{2} \leq \tau \\
% 0,\left\|\hat{\bf{g_{i}}}-\hat{\bf{g_{j}}}\right\|_{2}^{2}>\tau
% \end{array}\right.,
% \end{equation}
% where $\hat{\bf{g_{i}}}$ and $\hat{\bf{g_{j}}}$ represent the geometry information of $\hat{\bf{p_{i}}}$ and $\hat{\bf{p_{j}}}$, respectively. $\sigma$ is the variance of region vertices, and $\tau$ is the Euclidean distance threshold for clustering neighbors into the same region. In our experiments, we set $\tau=50$, and the relationship between the two parameters can be written as $\sigma^{2}=\tau^{2} / 2$.

\subsection{Geometry Density}
Based on the constructed local region, we exploit the geometry components to derivate geometry density features. To be specific, since the geometry information can be directly obtained from the local region, we separately compute the mean and standard deviation values of all three coordinates to serve as the geometry density features. For example, the mean and standard deviation of geometry information along $\rm{x}$-coordinate are calculated by:

\begin{equation}
\mu_{\bf{g}}^{\bf{x}}=\frac{\sum_{i=1}^{I} \tilde{\bf{g}}_{i}^{\bf{x}}}{I},
\end{equation}

\begin{equation}
\delta_{\bf{g}}^{\bf{x}}=\sqrt{\frac{\sum_{i=1}^{I}\left(\tilde{\bf{g}}_{i}^{\bf{x}}-\mu_{\bf{g}}^{\bf{x}}\right)^{2}}{I}},
\end{equation}
where $\tilde{\bf{g}}_{i}^{\bf{x}}$ and $I$ represent the $i-th$ geometry information and total number of graph points in the local region along $\rm{x}$-coordinate, respectively. Similar to $\rm{x}$-coordinate, we can also compute the mean and standard deviation of geometry information along $\rm{y}$-coordinate and $\rm{z}$-coordinate. With all three coordinates, we can obtain 6 dimensional geometry density features from the average operation of extracted features across various local regions.

\begin{figure*}[t]
	\centerline{\includegraphics[width=18.8cm]{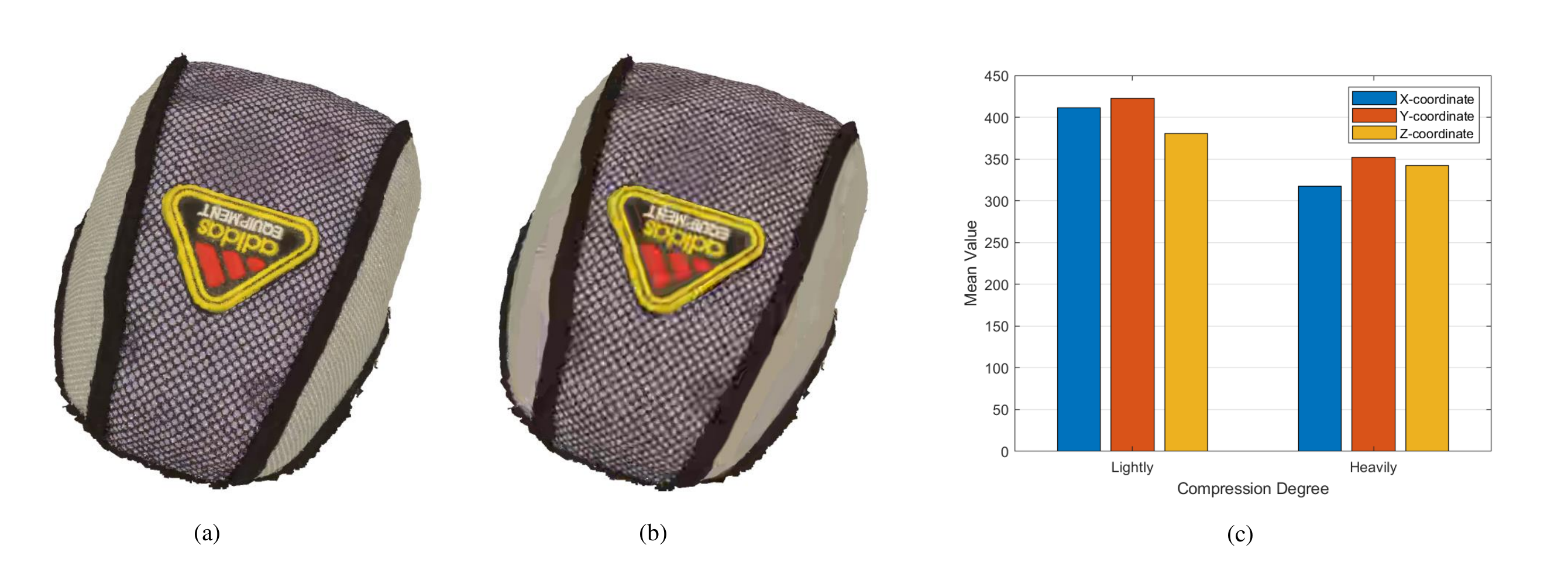}}
	\caption{Mean value changes for 3D point clouds at different distortion degrees. (a) Lightly compressed 3D point cloud, (b) heavily compressed 3D point cloud, (c) the mean value changes of (a,b).}
	\centering
	\label{figure4}
\end{figure*}

Since the standard deviation is induced from mean value, we provide the mean value changes for two 3D point clouds at different distortion degrees, as shown in Fig. \ref{figure4}. Here, the point clouds are distorted with light and heavy V-PCC compression artifacts. As can be seen from this figure, the histogram comparisons of the distorted 3D point clouds demonstrate that the mean value from all three coordinates can effectively reveal the quality variation. In addition, the effectiveness of these features is mainly because they can reflect the density of 3D point clouds.

\subsection{Color Naturalness}
As for 3D point clouds, color information is a significant attribute. Thus, we further extract color naturalness features from the constructed local region. By taking the human perception into account, we first convert the RGB components to YUV space. Then, to normalize the distributions, we apply the zero-phase component analysis followed by local mean subtraction and divisive normalization \cite{ruderman1994statistics} to the separate YUV channels of distorted local regions as:

\begin{equation}
\overline{\bf{\rho}}_{z}(y, u, v)=\frac{\tilde{\bf{\rho}}_{z}(y, u, v)-\mu(y, u, v)}{\delta(y, u, v)+\rm{C}}
\end{equation}
where $\tilde{\bf{\rho}}_{z}(y, u, v)$ is the YUV information of local regions after the whitening filter of zero-phase component analysis \cite{chen2017blind,zhou2021no}, and $\overline{\bf{\rho}}_{z}(y, u, v)$ represents the final normalized YUV channels. $\rm{C}=1$ is a small constant that avoids instabilities when the denominator tends to $0$. The local mean and standard deviation of YUV channels can be computed by:

\begin{equation}
\mu(y, u, v)=\sum_{l=-L}^{L} \sum_{t=-T}^{T} \omega_{l, t} \tilde{\bf{\rho}}_{z}(y, u, v),
\end{equation}

\begin{equation}
\delta(y, u, v)=\sqrt{\sum_{l=-L}^{L} \sum_{t=-T}^{T} \omega_{l, t}\left(\tilde{\bf{\rho}}_{z}(y, u, v)-\delta(y, u, v)\right)^{2}},
\end{equation}
where $\omega=\left\{\omega_{l, t} \mid l=-L, \ldots, L, t=-T, \ldots, T\right\}$ indicates a circularly symmetric Gaussian weighting function. Motivated by \cite{mittal2012no}, we set $L=T=3$.

\begin{figure}[t]
	\centerline{\includegraphics[width=9.8cm]{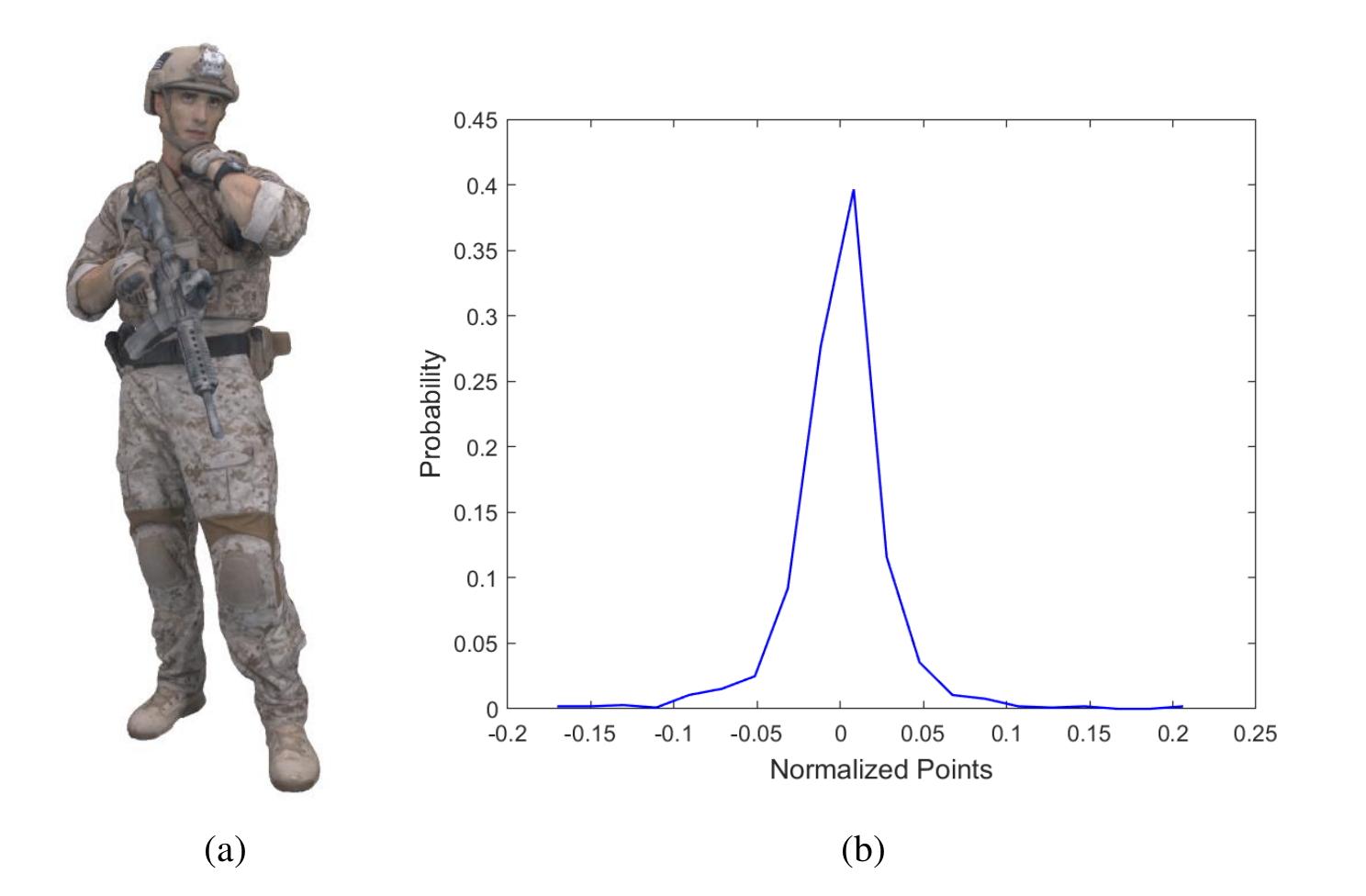}}
	\caption{An example of statistical probability distribution for normalized coefficients. (a) A distorted 3D point cloud, (b) the corresponding statistical probability distribution for normalized coefficients of (a).}
	\centering
	\label{figure5}
\end{figure}

In Fig. \ref{figure5}, we show an example of statistical probability distribution for normalized coefficients, where the 3D point cloud is distorted by downscaling artifacts. On the basis of its Gaussian-like appearance, we use the generalized Gaussian distribution (GGD) \cite{sharifi1995estimation} to capture the regularity of NSS, which is given by:

\begin{equation}
f\left(x ; \lambda, \varepsilon^{2}\right)=\frac{\lambda}{2 \eta \Gamma(1 / \lambda)} e^{-\left(\frac{|x|}{\eta}\right)^{\lambda}},
\end{equation}
where
\begin{equation}
\eta=\varepsilon \sqrt{\frac{\Gamma(1 / \lambda)}{\Gamma(3 / \lambda)}},
\end{equation}
and $\Gamma(\cdot)$ is the gamma function as follows:

\begin{equation}
\Gamma(a)=\int_{0}^{+\infty} x^{a-1} e^{-x} d x, a>0.
\end{equation}

In the GGD, $\lambda$ controls the shape of statistical probability distribution and $\varepsilon^{2}$ reflects the variance. For each color channel of the input distorted point cloud, we estimate 2 parameters $(\lambda, \varepsilon^{2})$ from the GGD fit by the moment matching-based approach \cite{sharifi1995estimation}. Additionally, we use four scales containing the original scale as well as the reduction by the factors of 2, 4, and 8. Totally, with YUV channels, we have 24 dimensional color naturalness features which are calculated by the average operation among all the local regions.

\subsection{Angular Consistency}
Another important attribute information is normal that indicates the orientation of points. To fully use the unique normal vectors, apart from regional pre-processing, they have also been verified to be correlated to visual quality \cite{tian2017geometric,alexiou2018point,viola2020reduced}. Therefore, we propose angular consistency features based on the normal information. Specifically, for each 3D point in the distorted point clouds, we first estimate the normal vector $\vec{\chi}_{i}$ of the point by using multiple neighboring points to fit a local plane for determining the normal vector. It should be noted that the number of neighboring points for normal vector estimation is validated in the experiments. Then, we adopt the $k$-nearest algorithm to choose the set of neighbors for each normal vector. In such case, assume that the normal vector of each neighbor point is denoted by $\vec{\chi}_{j}$, we compute the cosine similarity between $\vec{\chi}_{i}$ and $\vec{\chi}_{j}$ by:

\begin{equation}
\Omega=\cos (\zeta)=\frac{\vec{\chi}_{i} \cdot \vec{\chi}_{j}}{\left\|\vec{\chi}_{i}\right\| *\left\|\vec{\chi}_{j}\right\|},
\end{equation}
where $\Omega\in [-1, 1]$ and $\zeta$ is the angular between the two normal vectors. Based on this, the inverse cosine of the obtained $\Omega$ is calculated as:

\begin{equation}
\zeta^{\prime}=\arccos (|\Omega|),
\end{equation}
where $\zeta^{\prime}\in [0, \pi/2]$ and thus the ultimate angular similarity can be computed in the range of $[0, 1]$ as follows:

\begin{equation}
\psi=1-\frac{2 \zeta^{\prime}}{\pi}.
\end{equation}

\begin{figure}[t]
	\centerline{\includegraphics[width=9.8cm]{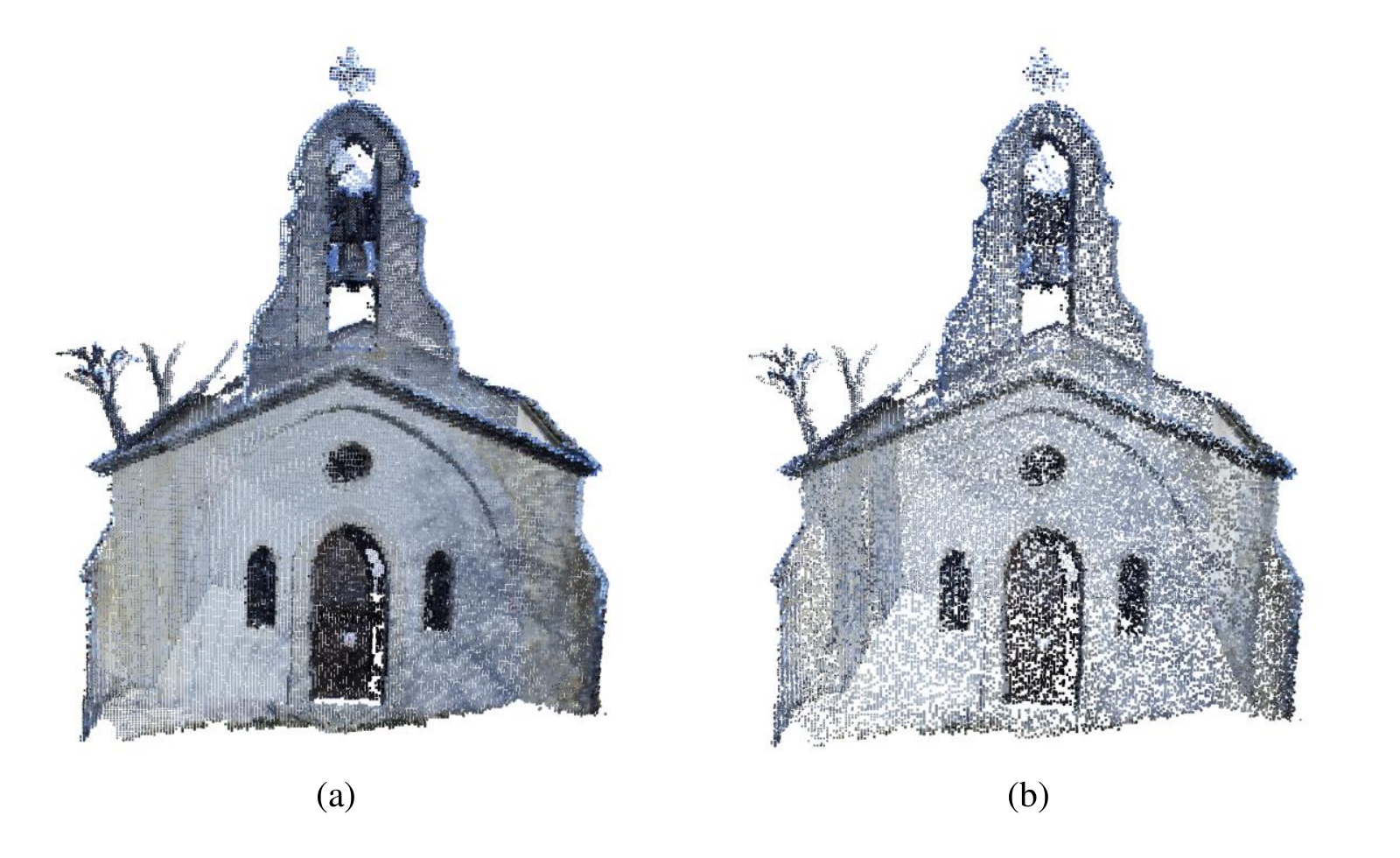}}
	\caption{Demonstration of two scales for distorted 3D point clouds. (a) A distorted 3D point cloud, (b) the downsampled scale of (a).}
	\centering
	\label{figure6}
\end{figure}

From the above-mentioned computation, we obtain the angular similarity matrix for every distorted point cloud. On one hand, we conduct an average operation for the point number dimension. On the other hand, 5 kinds of statistics are calculated for the $k$-nearest neighbor dimension, where the first two statistical features are the mean $\mu$ and standard deviation $\delta$. Furthermore, the skewness and kurtosis are computed as:

\begin{equation}
S=E\left[ \left( \frac{\psi_k - \mu}{\delta}\right)^3 \right],
\end{equation}

\begin{equation}
K=E\left[ \left( \frac{\psi_k - \mu}{\delta}\right)^4 \right],
\end{equation}
where $E[\cdot]$ denotes the expectation operator, and $\psi_k$ is the angular similarity value. Except for these moments, as suggested in \cite{fang2014no}, the entropy is also obtained by:

\begin{equation}
H=-\sum_{k} p(\psi_k) \rm{log} p(\psi_k),
\end{equation}
where $p(\psi_k)$ is the probability of $\psi_k$. These statistics of angular similarity matrix can be regarded as the angular consistency of normal vectors.

To extract the angular consistency features, since the HVS perception is hierarchical, here we employ more scales for better feature representations. As illustrated in Fig. \ref{figure6}, two scales of PCL compressed point clouds are shown, where figures (a) and (b) have $132,518$ points and $66,259$ points, respectively. From the two scales, we have 10 dimensional angular consistency features, which are used for quantifying visual distortions of 3D point clouds.

\begin{table*}[t]
\begin{center}
\captionsetup{justification=centering}
\caption{\textsc{Detailed Information of Subjective Point Cloud Quality Databases.}}
\label{table1}
\scalebox{0.95}{
\begin{tabular}{|c|c|c|c|c|}
\hline
Databases & Reference Number & Distortion Number & Distortion Types & MOS Range \\ \hline
Waterloo & 20 & 660 & Downsampling, Gaussian noise, G-PCC, V-PCC & [1, 100] \\
M-PCCD & 8 & 232 & Octree-Lifting, Octree-RAHT, TriSoup-Lifting, TriSoup-RAHT, V-PCC & [1, 5] \\
SJTU & 6 & 144 & Octree-based compression, Color noise, Downscaling, Geometry Gaussian noise & [1, 10] \\
IRPC & 6 & 54 & PCL, G-PCC, V-PCC & [1, 5] \\ \hline
\end{tabular}}
\end{center}
\end{table*}

\begin{table*}[ht]
	\begin{center}
		\captionsetup{justification=centering}
		\caption{\textsc{Performance Evaluation of Different Point Cloud Contents.}}
		\label{table2}
		\setlength{\tabcolsep}{8mm}{
			\scalebox{0.79}{
                \begin{tabular}{|c|c|c|c|c|c|}
				\hline
				Point Cloud Contents & Method Types & Method Names & SROCC & KROCC & PLCC \\ \hline
				    \multirow{17}{*}{Banana} & \multirow{14}{*}{FR} & $PSNR_{MSE, p2po}$ \cite{mekuria2016evaluation} &0.64 &0.47 &0.73 \\
                    & & $PSNR_{HF, p2po}$ \cite{mekuria2016evaluation} &0.07 &0.02 &0.35 \\
                    & & $PSNR_{MSE, p2pl}$ \cite{tian2017geometric} &0.54 &0.40 &0.57 \\
                    & & $PSNR_{HF, p2pl}$ \cite{tian2017geometric} &0.08 &0.04 &0.35 \\
                    & & $PSNR_{Y}$ \cite{mekuria2017performance} &0.62 &0.47 &0.72 \\
                    & & $AS_{Mean}$ \cite{alexiou2018point} &0.39 &0.30 &0.40 \\
                    & & $AS_{RMS}$ \cite{alexiou2018point} &0.34 &0.26 &0.39 \\
                    & & $AS_{MSE}$ \cite{alexiou2018point} &0.34 &0.26 &0.39 \\
                    & & $SSIM_{projected}$ \cite{wang2004image} &0.77 &0.60 &0.72 \\
                    & & $MS$-$SSIM_{projected}$ \cite{wang2003multiscale} &0.82 &0.65 &0.80 \\
                    & & $VIFP_{projected}$ \cite{sheikh2006image} &0.82 &0.64 &0.81 \\
                    & & $GraphSIM$ \cite{yang2020inferring} &0.46 &0.37 &0.53 \\
                    & & $PCQM$ \cite{meynet2020pcqm} &0.74 &0.56 &0.64 \\
                    & & $PointSSIM$ \cite{alexiou2020towards} &0.18 &0.12 &0.10 \\ \cline{2-6}
                    & \multirow{1}{*}{RR} & $PCMRR$ \cite{viola2020reduced} &0.47 &0.38 &0.42 \\ \cline{2-6}
                    & \multirow{2}{*}{NR} & $PQA$-$Net$ \cite{liu2021pqa} &0.52 &0.39 &0.53 \\
                    & & \textbf{Proposed SGR (SVR)} &0.68 &0.51 &0.70 \\
                    & & \textbf{Proposed SGR (RFR)} &\textbf{0.86} &\textbf{0.68} &\textbf{0.83} \\ \hline
                    \multirow{17}{*}{Cauliflower} & \multirow{14}{*}{FR} & $PSNR_{MSE, p2po}$ \cite{mekuria2016evaluation} &0.49 &0.35 &0.55 \\
                    & & $PSNR_{HF, p2po}$ \cite{mekuria2016evaluation} &0.16 &0.11 &0.09 \\
                    & & $PSNR_{MSE, p2pl}$ \cite{tian2017geometric} &0.30 &0.23 &0.40 \\
                    & & $PSNR_{HF, p2pl}$ \cite{tian2017geometric} &0.25 &0.18 &0.33 \\
                    & & $PSNR_{Y}$ \cite{mekuria2017performance} &0.54 &0.40 &0.56 \\
                    & & $AS_{Mean}$ \cite{alexiou2018point} &0.24 &0.18 &0.37 \\
                    & & $AS_{RMS}$ \cite{alexiou2018point} &0.24 &0.18 &0.37 \\
                    & & $AS_{MSE}$ \cite{alexiou2018point} &0.24 &0.18 &0.37 \\
                    & & $SSIM_{projected}$ \cite{wang2004image} &0.77 &0.61 &0.80 \\
                    & & $MS$-$SSIM_{projected}$ \cite{wang2003multiscale} &0.74 &0.58 &0.76 \\
                    & & $VIFP_{projected}$ \cite{sheikh2006image} &0.75 &0.59 &0.80 \\
                    & & $GraphSIM$ \cite{yang2020inferring} &0.59 &0.43 &0.61 \\
                    & & $PCQM$ \cite{meynet2020pcqm} &0.69 &0.52 &0.67 \\
                    & & $PointSSIM$ \cite{alexiou2020towards} &0.21 &0.19 &0.36 \\ \cline{2-6}
                    & \multirow{1}{*}{RR} & $PCMRR$ \cite{viola2020reduced} &0.29 &0.20 &0.30 \\ \cline{2-6}
                    & \multirow{2}{*}{NR} & $PQA$-$Net$ \cite{liu2021pqa} &0.69 &\textbf{0.52} &0.70 \\
                    & & \textbf{Proposed SGR (SVR)} &0.68 &0.50 &0.70 \\
                    & & \textbf{Proposed SGR (RFR)} &\textbf{0.70} &\textbf{0.52} &\textbf{0.72} \\ \hline
                    \multirow{17}{*}{Mushroom} & \multirow{14}{*}{FR} & $PSNR_{MSE, p2po}$ \cite{mekuria2016evaluation} &0.63 &0.48 &0.66 \\
                    & & $PSNR_{HF, p2po}$ \cite{mekuria2016evaluation} &0.26 &0.20 &0.48 \\
                    & & $PSNR_{MSE, p2pl}$ \cite{tian2017geometric} &0.47 &0.37 &0.55 \\
                    & & $PSNR_{HF, p2pl}$ \cite{tian2017geometric} &0.22 &0.16 &0.45 \\
                    & & $PSNR_{Y}$ \cite{mekuria2017performance} &0.60 &0.44 &0.79 \\
                    & & $AS_{Mean}$ \cite{alexiou2018point} &0.26 &0.18 &0.39 \\
                    & & $AS_{RMS}$ \cite{alexiou2018point} &0.26 &0.18 &0.39 \\
                    & & $AS_{MSE}$ \cite{alexiou2018point} &0.26 &0.18 &0.39 \\
                    & & $SSIM_{projected}$ \cite{wang2004image} &0.73 &0.57 &0.85 \\
                    & & $MS$-$SSIM_{projected}$ \cite{wang2003multiscale} &0.89 &0.73 &0.88 \\
                    & & $VIFP_{projected}$ \cite{sheikh2006image} &0.90 &0.76 &0.90 \\
                    & & $GraphSIM$ \cite{yang2020inferring} &0.65 &0.50 &0.68 \\
                    & & $PCQM$ \cite{meynet2020pcqm} &0.76 &0.60 &0.71 \\
                    & & $PointSSIM$ \cite{alexiou2020towards} &0.33 &0.26 &0.36 \\ \cline{2-6}
                    & \multirow{1}{*}{RR} & $PCMRR$ \cite{viola2020reduced} &0.18 &0.14 &0.19 \\ \cline{2-6}
                    & \multirow{2}{*}{NR} & $PQA$-$Net$ \cite{liu2021pqa} &0.71 &0.56 &0.77 \\
                    & & \textbf{Proposed SGR (SVR)} &0.80 &\textbf{0.66} &0.85 \\
                    & & \textbf{Proposed SGR (RFR)} &\textbf{0.82} &0.65 &\textbf{0.86} \\ \hline
                    \multirow{17}{*}{Overall} & \multirow{14}{*}{FR} & $PSNR_{MSE, p2po}$ \cite{mekuria2016evaluation} &0.48 &0.33 &0.50 \\
                    & & $PSNR_{HF, p2po}$ \cite{mekuria2016evaluation} &0.16 &0.11 &0.34 \\
                    & & $PSNR_{MSE, p2pl}$ \cite{tian2017geometric} &0.36 &0.25 &0.37 \\
                    & & $PSNR_{HF, p2pl}$ \cite{tian2017geometric} &0.20 &0.14 &0.27 \\
                    & & $PSNR_{Y}$ \cite{mekuria2017performance} &0.53 &0.37 &0.56 \\
                    & & $AS_{Mean}$ \cite{alexiou2018point} &0.24 &0.16 &0.24 \\
                    & & $AS_{RMS}$ \cite{alexiou2018point} &0.21 &0.14 &0.23 \\
                    & & $AS_{MSE}$ \cite{alexiou2018point} &0.21 &0.14 &0.23 \\
                    & & $SSIM_{projected}$ \cite{wang2004image} &0.58 &0.42 &0.59 \\
                    & & $MS$-$SSIM_{projected}$ \cite{wang2003multiscale} &0.60 &0.44 &0.62 \\
                    & & $VIFP_{projected}$ \cite{sheikh2006image} &0.82 &0.63 &0.82 \\
                    & & $GraphSIM$ \cite{yang2020inferring} &0.46 &0.32 &0.47 \\
                    & & $PCQM$ \cite{meynet2020pcqm} &0.71 &0.52 &0.65 \\
                    & & $PointSSIM$ \cite{alexiou2020towards} &0.18 &0.14 &0.14 \\ \cline{2-6}
                    & \multirow{1}{*}{RR} & $PCMRR$ \cite{viola2020reduced} &0.26 &0.19 &0.29 \\ \cline{2-6}
                    & \multirow{2}{*}{NR} & $PQA$-$Net$ \cite{liu2021pqa} &0.69 &0.51 &0.70 \\
                    & & \textbf{Proposed SGR (SVR)} &\textbf{0.73} &\textbf{0.55} &0.74 \\
                    & & \textbf{Proposed SGR (RFR)} &\textbf{0.73} &\textbf{0.55} &\textbf{0.75} \\ \hline		
		\end{tabular}}}
	\end{center}
\end{table*}

\begin{table*}[ht]
	\begin{center}
		\captionsetup{justification=centering}
		\caption{\textsc{Performance Evaluation of Various Distortion Types.}}
		\label{table3}
		\setlength{\tabcolsep}{8mm}{
			\scalebox{0.79}{
                \begin{tabular}{|c|c|c|c|c|c|}
				\hline
				Distortion Types & Method Types & Method Names & SROCC & KROCC & PLCC \\ \hline
				    \multirow{17}{*}{Downsampling} & \multirow{14}{*}{FR} & $PSNR_{MSE, p2po}$ \cite{mekuria2016evaluation} &0.86 &0.67 &0.96 \\
                    & & $PSNR_{HF, p2po}$ \cite{mekuria2016evaluation} &0.85 &0.64 &0.97 \\
            		& & $PSNR_{MSE, p2pl}$ \cite{tian2017geometric} &0.76 &0.52 &0.85 \\
            		& & $PSNR_{HF, p2pl}$ \cite{tian2017geometric} &0.84 &0.61 &0.97 \\
            		& & $PSNR_{Y}$ \cite{mekuria2017performance} &0.69 &0.55 &0.70 \\
            		& & $AS_{Mean}$ \cite{alexiou2018point} &0.77 &0.52 &0.96 \\
            		& & $AS_{RMS}$ \cite{alexiou2018point} &0.77 &0.52 &0.96 \\
            		& & $AS_{MSE}$ \cite{alexiou2018point} &0.77 &0.52 &0.96 \\
            		& & $SSIM_{projected}$ \cite{wang2004image} &0.87 &0.73 &0.91 \\
            		& & $MS$-$SSIM_{projected}$ \cite{wang2003multiscale} &0.83 &0.64 &0.89 \\
            		& & $VIFP_{projected}$ \cite{sheikh2006image} &0.91 &0.76 &0.98 \\
            		& & $GraphSIM$ \cite{yang2020inferring} &0.79 &0.64 &0.97 \\
            		& & $PCQM$ \cite{meynet2020pcqm} &0.89 &0.73 &0.85 \\
            		& & $PointSSIM$ \cite{alexiou2020towards} &0.91 &0.76 &0.97 \\ \cline{2-6}
            	    & \multirow{1}{*}{RR} & $PCMRR$ \cite{viola2020reduced} &0.88 &0.70 &0.89 \\ \cline{2-6}
            	    & \multirow{2}{*}{NR} & $PQA$-$Net$ \cite{liu2021pqa} &0.80 &0.64 &\textbf{0.97} \\
            	    & & \textbf{Proposed SGR (SVR)} &0.82 &0.61 &0.95 \\
                    & & \textbf{Proposed SGR (RFR)} &\textbf{0.84} &\textbf{0.67} &\textbf{0.97} \\ \hline	
                    \multirow{17}{*}{Gaussian Noise} & \multirow{14}{*}{FR} & $PSNR_{MSE, p2po}$ \cite{mekuria2016evaluation} &0.63 &0.44 &0.67 \\
                    & & $PSNR_{HF, p2po}$ \cite{mekuria2016evaluation} &0.63 &0.45 &0.67 \\
            		& & $PSNR_{MSE, p2pl}$ \cite{tian2017geometric} &0.62 &0.43 &0.67 \\
            		& & $PSNR_{HF, p2pl}$ \cite{tian2017geometric} &0.63 &0.45 &0.67 \\
            		& & $PSNR_{Y}$ \cite{mekuria2017performance} &0.84 &0.68 &0.90 \\
            		& & $AS_{Mean}$ \cite{alexiou2018point} &0.68 &0.49 &0.71 \\
            		& & $AS_{RMS}$ \cite{alexiou2018point} &0.67 &0.49 &0.71 \\
            		& & $AS_{MSE}$ \cite{alexiou2018point} &0.67 &0.49 &0.70 \\
            		& & $SSIM_{projected}$ \cite{wang2004image} &0.66 &0.48 &0.84 \\
            		& & $MS$-$SSIM_{projected}$ \cite{wang2003multiscale} &0.66 &0.51 &0.85 \\
            		& & $VIFP_{projected}$ \cite{sheikh2006image} &0.81 &0.66 &0.86 \\
            		& & $GraphSIM$ \cite{yang2020inferring} &0.71 &0.57 &0.72 \\
            		& & $PCQM$ \cite{meynet2020pcqm} &0.87 &0.70 &0.89 \\
            		& & $PointSSIM$ \cite{alexiou2020towards} &0.63 &0.46 &0.70 \\ \cline{2-6}
            	    & \multirow{1}{*}{RR} & $PCMRR$ \cite{viola2020reduced} &0.88 &0.73 &0.89 \\ \cline{2-6}
            	    & \multirow{2}{*}{NR} & $PQA$-$Net$ \cite{liu2021pqa} &0.64 &0.44 &0.75 \\
            		& & \textbf{Proposed SGR (SVR)} &\textbf{0.91} &\textbf{0.74} &\textbf{0.94} \\
                    & & \textbf{Proposed SGR (RFR)} &0.85 &0.68 &0.92 \\ \hline			
                    \multirow{17}{*}{G-PCC} & \multirow{14}{*}{FR} & $PSNR_{MSE, p2po}$ \cite{mekuria2016evaluation} &0.39 &0.29 &0.41 \\
                    & & $PSNR_{HF, p2po}$ \cite{mekuria2016evaluation} &0.42 &0.30 &0.40 \\
            		& & $PSNR_{MSE, p2pl}$ \cite{tian2017geometric} &0.42 &0.30 &0.42 \\
            		& & $PSNR_{HF, p2pl}$ \cite{tian2017geometric} &0.34 &0.23 &0.34 \\
            		& & $PSNR_{Y}$ \cite{mekuria2017performance} &0.73 &0.55 &0.75 \\
            		& & $AS_{Mean}$ \cite{alexiou2018point}	&0.03 &0.08 &0.13 \\
            		& & $AS_{RMS}$ \cite{alexiou2018point} &0.03 &0.03 &0.13 \\
            		& & $AS_{MSE}$ \cite{alexiou2018point} &0.03 &0.03 &0.13 \\
            		& & $SSIM_{projected}$ \cite{wang2004image} &0.63 &0.48 &0.63 \\
            		& & $MS$-$SSIM_{projected}$ \cite{wang2003multiscale} &0.66 &0.51 &0.67 \\
            		& & $VIFP_{projected}$ \cite{sheikh2006image} &0.83 &0.65 &0.84 \\
            		& & $GraphSIM$ \cite{yang2020inferring} &0.61 &0.44 &0.70 \\
            		& & $PCQM$ \cite{meynet2020pcqm} &0.85 &0.69 &0.72 \\
            		& & $PointSSIM$ \cite{alexiou2020towards} &0.77 &0.60 &0.78 \\ \cline{2-6}
            	    & \multirow{1}{*}{RR} & $PCMRR$ \cite{viola2020reduced} &0.19 &0.13 &0.20 \\ \cline{2-6}
            	    & \multirow{2}{*}{NR} & $PQA$-$Net$ \cite{liu2021pqa} &0.67 &0.51 &0.68 \\
            		& & \textbf{Proposed SGR (SVR)} &\textbf{0.70} &\textbf{0.52} &\textbf{0.73} \\
                    & & \textbf{Proposed SGR (RFR)} &0.69 &0.50 &0.72 \\ \hline
                    \multirow{17}{*}{V-PCC} & \multirow{14}{*}{FR} & $PSNR_{MSE, p2po}$ \cite{mekuria2016evaluation} &0.42 &0.30 &0.48 \\
                    & & $PSNR_{HF, p2po}$ \cite{mekuria2016evaluation} &0.27 &0.19 &0.35 \\
            		& & $PSNR_{MSE, p2pl}$ \cite{tian2017geometric} &0.46 &0.32 &0.48 \\
            		& & $PSNR_{HF, p2pl}$ \cite{tian2017geometric} &0.43 &0.32 &0.52 \\
            		& & $PSNR_{Y}$ \cite{mekuria2017performance} &0.32 &0.25 &0.48 \\
            		& & $AS_{Mean}$ \cite{alexiou2018point} &0.53 &0.35 &0.66 \\
            		& & $AS_{RMS}$ \cite{alexiou2018point} &0.49 &0.32 &0.60 \\
            		& & $AS_{MSE}$ \cite{alexiou2018point} &0.49 &0.32 &0.60 \\
            		& & $SSIM_{projected}$ \cite{wang2004image} &0.35 &0.29 &0.44 \\
            		& & $MS$-$SSIM_{projected}$ \cite{wang2003multiscale} &0.38 &0.34 &0.50 \\
            		& & $VIFP_{projected}$ \cite{sheikh2006image} &0.90 &0.76 &0.91 \\
            		& & $GraphSIM$ \cite{yang2020inferring} &0.32 &0.27 &0.61 \\
            		& & $PCQM$ \cite{meynet2020pcqm} &0.59 &0.45 &0.59 \\
            		& & $PointSSIM$ \cite{alexiou2020towards}	&0.48 &0.34 &0.51 \\ \cline{2-6}
            	    & \multirow{1}{*}{RR} & $PCMRR$ \cite{viola2020reduced} &0.55 &0.42 &0.50 \\ \cline{2-6}
            	    & \multirow{2}{*}{NR} & $PQA$-$Net$ \cite{liu2021pqa} &0.45 &0.30 &0.60 \\
            		& & \textbf{Proposed SGR (SVR)} &\textbf{0.75} &\textbf{0.57} &\textbf{0.79} \\
                    & & \textbf{Proposed SGR (RFR)} &0.72 &0.53 &0.76 \\ \hline			
		\end{tabular}}}
	\end{center}
\end{table*}

\subsection{Quality Regression}
After the regional pre-processing and quality-aware feature extraction, a quality regression module is applied to map the relevant features onto final quality scores. Many regressors can be used for this purpose, among whom we select the well-known support vector regression (SVR) and random forest regression (RFR) due to their capacity and popularity in quality assessment problems \cite{gu2014using,shi2019no,dendi2020no,zhou2020tensor}. Especially, they have also been widely used for the spatial-domain and transform-domain NSS in the 2D image quality evaluation task \cite{moorthy2011blind,saad2012blind,pei2015image,freitas2018no}.

In our framework, given a test 3D point cloud, all features from geometry density, color naturalness and angular consistency aspects are extracted and concatenated to a feature vector. Before predicting the quality result of the test 3D point cloud, we use the labeled point clouds to train the regressor. With the trained regressor, we can predict the perceptual quality of any input test 3D point cloud. The LIBSVM \cite{chang2011libsvm} is utilized to implement SVR with a radial basis function kernel, together with the RFR proposed in \cite{breiman2001random,criminisi2011decision} are used for our experiments. Additionally, we follow the common regressor parameter settings that have been used in the training of mainstream quality assessment models.

\section{Validation of Proposed Method}

\subsection{Evaluation Databases and Criteria}
We conduct experiments to validate our proposed SGR on four publicly available subjective point cloud quality databases, including Waterloo \cite{su2019perceptual,liu2022perceptual}, M-PCCD \cite{alexiou2019comprehensive}, SJTU \cite{yang2020predicting} and IRPC \cite{javaheri2020point} databases. The detailed information of these databases is shown in TABLE \ref{table1}. Specifically, followed by the recent NR quality assessment work \cite{liu2021pqa}, we first use the selected Waterloo database to compare the proposed SGR with a variety of state-of-the-art FR, RR and NR quality assessment methods for 3D point clouds. This database contains 20 original reference point cloud contents with different geometry and texture complexities. Four distortion types are considered to generate the corresponding quality-degraded point clouds, which include 60 downsampling distorted point clouds, 180 distorted point clouds with Gaussian noise, 320 G-PCC (T) compressed point clouds, and 180 V-PCC compressed point clouds. Moreover, the adopted SJTU database consists of 6 pristine point cloud contents, i.e. longdress, loot, redandblack, shiva, soldier, and statue. To produce distorted point clouds, the Octree-based compression, color noise, downsampling, and geometry Gaussian noise are introduced to the original point clouds. As for M-PCCD and IRPC databases, the numbers of original point cloud contents are 8 and 6, respectively. By involving several distortion types, 232 and 54 distorted 3D point clouds are generated in the two databases.

Apart from the reference and distorted point clouds, each database also provides subjective quality labels in the form of mean opinion score (MOS), which are rated by viewers with several display modes, such as direct point cloud format and converted video format, etc. Details can be referred to the literature \cite{su2019perceptual,liu2022perceptual,alexiou2019comprehensive,yang2020predicting,javaheri2020point}. Note that only the 3D point clouds in the IRPC database have normal vectors.

\begin{figure*}[t]
	\centerline{\includegraphics[width=18.7cm]{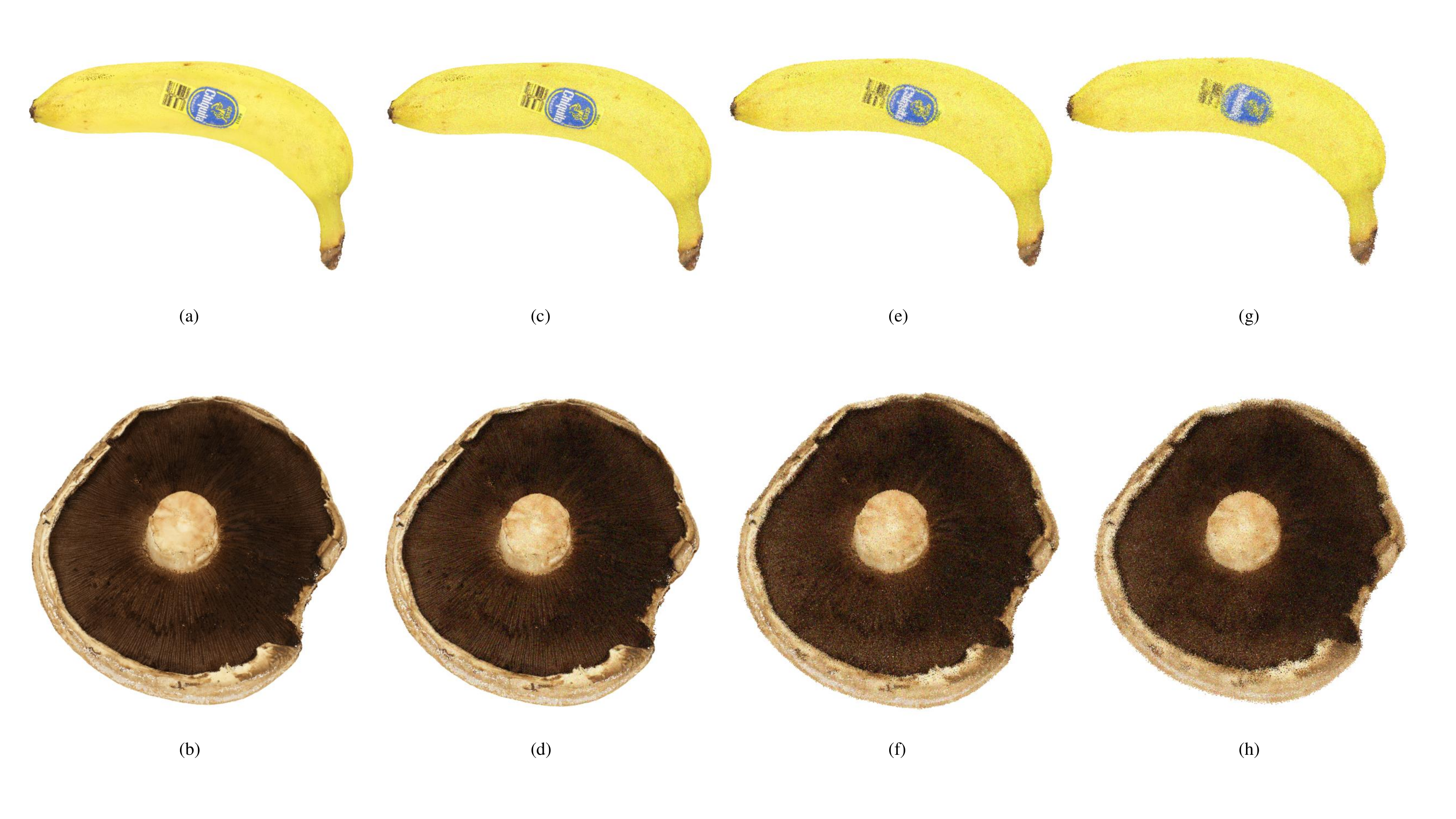}}
	\caption{Examples of perceptual quality prediction. (a-b) Original reference point clouds with different contents, and the corresponding distortion degree increases from left to right. (c) SGR(SVR)=74.4780 / SGR(RFR)=61.9948, (d) SGR(SVR)=73.5027 / SGR(RFR)=68.9962, (e) SGR(SVR)=49.7316 / SGR(RFR)=46.3371, (f) SGR(SVR)=43.2950 / SGR(RFR)=44.6210, (g) SGR(SVR)=38.7809 / SGR(RFR)=36.8399, (h) SGR(SVR)=36.6015 / SGR(RFR)=35.2080. Higher MOS or SGR represents better visual quality for point clouds.}
	\centering
	\label{figure7}
\end{figure*}

To validate the proposed SGR and compare it with other state-of-the-arts, we adopt three commonly-used evaluation criteria in IQA field, including Spearman Rank-Order Correlation Coefficient (SROCC), Kendall Rank-Order Correlation Coefficient (KROCC), and Pearson Linear Correlation Coefficient (PLCC). Among these evaluation criteria, the SROCC is usually used to measure prediction monotonicity, while KROCC can be applied to evaluate the ordinal association between two measured quantities. Moreover, the PLCC can be used to evaluate prediction accuracy. It should be noted that higher correlation coefficients indicate better performance for objective quality models.

Besides, before computing the PLCC for different objective quality assessment approaches, a five-parameter logistic nonlinear fitting function \cite{rohaly2000video} is adopted to map the predicted quality scores into a common scale as:

\begin{equation}
\rm{g(\gamma) = {\beta _1}(\frac{1}{2} - \frac{1}{{1 + {e^{({\beta _2}(\gamma - {\beta _3}))}}}}) + {\beta _4}\gamma + {\beta _5}},
\end{equation}
where $\rm{({\beta _1}...{\beta _5})}$ are five parameters to be fitted. $\rm{\gamma}$ represents the raw objective score produced by objective quality models and $\rm{g(\gamma)}$ denotes the regressed score after the nonlinear mapping.

\subsection{Performance of Objective Models}
In order to validate the performance of our proposed SGR, we compare the proposed SGR with state-of-the-art FR, RR and NR models on the largest subjective point cloud quality database, i.e. the Waterloo database \cite{su2019perceptual,liu2022perceptual}. These include 14 FR methods, where both point-based and projection-based approaches are compared. Among them, classical point-based metrics contain $PSNR_{MSE, p2po}$, $PSNR_{HF, p2po}$, $PSNR_{MSE, p2pl}$, $PSNR_{HF, p2pl}$, $PSNR_{Y}$, $AS_{Mean}$, $AS_{RMS}$, $AS_{MSE}$, $GraphSIM$, $PCQM$, and $PointSSIM$. The projection-based FR metrics involve $SSIM_{projected}$, $MS$-$SSIM_{projected}$, and $VIFP_{projected}$. Representative RR and NR methods are also taken into consideration, such as $PCMRR$ and $PQA$-$Net$.

By following \cite{liu2021pqa}, we conduct the comparison for various visual contents and distortion types. The evaluation results of different point cloud contents are reported in TABLE \ref{table2}. For space convenience, we show several contents and the overall performance. We can observe that our proposed SGR achieves competitive results for all the test visual contents. Besides, by using both the SVR and RFR regressors, our framework delivers promising mapping results, demonstrating that the proposed SGR does not rely on specific regressor. More importantly, our proposed SGR is superior to the $PQA$-$Net$ regarding the overall performance. It should be noted that the $PQA$-$Net$ is a deep learning-based NR quality assessment model for 3D point clouds. In addition, beyond the performance results for individual visual contents, we show the performance comparisons of various distortion types in TABLE \ref{table3}. Again, our proposed SGR outperforms the $PQA$-$Net$ for all distortion types, especially for Gaussian noise and V-PCC compression. Therefore, both the two tables demonstrate the superiority of our SGR algorithm.

\subsection{Visualization Comparisons}
In addition to quantitative performance results, we compare the visualized 3D point clouds with the predicted scores by our proposed SGR. Both SVR and RFR predicted quality results are computed.

We show some examples of perceptual quality prediction in Fig. \ref{figure7}. Each row represents one visual content. That is, the original reference contents are banana and mushroom, which contain 807,184 and 1144,603 points, respectively. By introducing the Gaussian noise, the reference point cloud would be distorted with various degrees. From the trends of SGR (SVR) and SGR (RFR) values, we can see that the proposed SGR method successfully distinguishes distorted point cloud data of different qualities. Therefore, the proposed model can effectively evaluate the perceptual quality of distorted 3D point clouds. In addition, for point clouds with relatively lower quality, i.e. figures (e-f) and (g-h), the prediction results from two models are more consistent. This may be because compared to low-quality point clouds, it is harder to predict the perceptual quality of high-quality point cloud data.

\subsection{Validity on Other Subjective Databases}
Apart from the largest Waterloo database, we further test the performance of the proposed SGR on other subjective databases, including M-PCCD \cite{alexiou2019comprehensive}, SJTU \cite{yang2020predicting}, and IRPC \cite{javaheri2020point}. Note that only the distorted 3D point clouds with individual/single distortion types are tested.

In the experiments, each database is randomly divided into training and testing sets with $80\%$-$20\%$ splitting. TABLE \ref{table4} provides the comparison results, where the state-of-the-art NR model $PQA$-$Net$ is compared with our proposed SGR. From this table, we can find that the proposed SGR performs better than the $PQA$-$Net$, which demonstrates the effectiveness of our SGR in general.

\begin{table}[t]
\begin{center}
\captionsetup{justification=centering}
\caption{\textsc{Performance Comparison on Other Subjective Databases.}}
\label{table4}
\scalebox{0.76}{
\begin{tabular}{|c|cc|cc|cc|}
\hline
Databases & \multicolumn{2}{c|}{M-PCCD} & \multicolumn{2}{c|}{SJTU} & \multicolumn{2}{c|}{IRPC} \\ \hline
Methods & SROCC & PLCC & SROCC & PLCC & SROCC & PLCC \\ \hline
$PQA$-$Net$ & 0.60 & 0.65 & 0.82 & 0.85 & 0.40 & 0.58 \\ \hline
\textbf{Proposed SGR (SVR)} & 0.80 & 0.87 & \textbf{0.89} & \textbf{0.93} & 0.67 & \textbf{0.89} \\
\textbf{Proposed SGR (RFR)} & \textbf{0.91} & \textbf{0.92} & 0.84 & 0.89 & \textbf{0.68} & \textbf{0.89} \\ \hline
\end{tabular}}
\end{center}
\end{table}

\begin{table}[t]
\begin{center}
\captionsetup{justification=centering}
\caption{\textsc{Performance Results of Individual Component on SJTU database.}}
\label{table5}
\scalebox{0.90}{
\begin{tabular}{|c|c|c|c|}
\hline
Methods & SROCC & KROCC & PLCC \\ \hline
Geometry Density & 0.58 & 0.41 & 0.70 \\ \hline
Color Naturalness & 0.54 & 0.40 & 0.72 \\ \hline
Geometry Density + Color Naturalness & 0.78 & 0.61 & 0.85 \\ \hline
Angular Consistency & 0.72 & 0.55 & 0.84 \\ \hline
\textbf{Proposed SGR (SVR)} & \textbf{0.89} & \textbf{0.72} & \textbf{0.93} \\ \hline
\end{tabular}}
\end{center}
\end{table}

\begin{table}[t]
\begin{center}
\captionsetup{justification=centering}
\caption{\textsc{Performance Results of Keypoint Resampling on SJTU database.}}
\label{table6}
\scalebox{1.0}{
\begin{tabular}{|c|c|c|c|}
\hline
Methods & SROCC & KROCC & PLCC \\ \hline
Color & 0.85 & 0.69 & 0.91 \\ \hline
Geometry & 0.86 & 0.69 & 0.91 \\ \hline
\textbf{Proposed SGR (SVR)} & \textbf{0.89} & \textbf{0.72} & \textbf{0.93} \\ \hline
\end{tabular}}
\end{center}
\end{table}

\begin{figure*}[t]
	\centerline{\includegraphics[width=19.8cm]{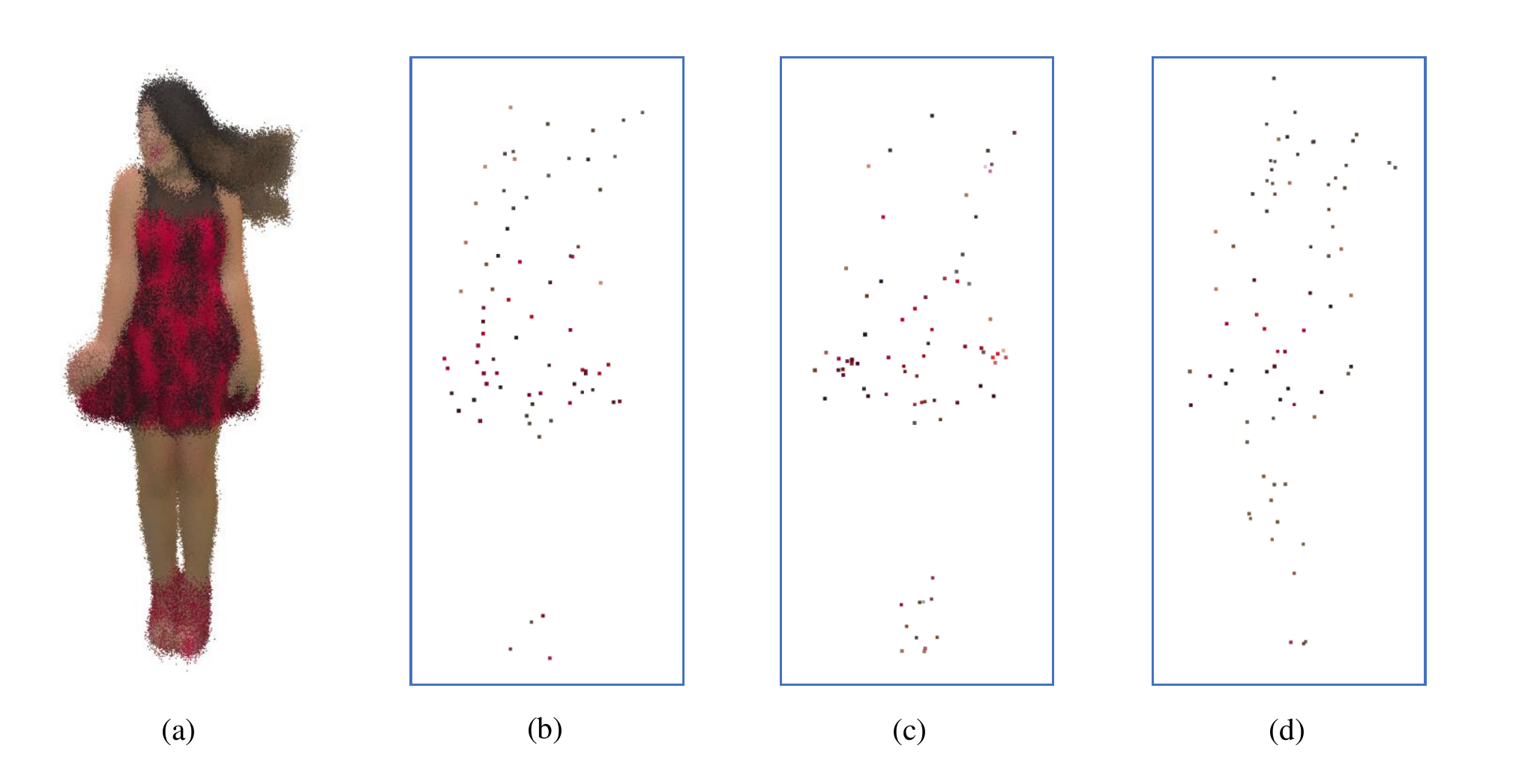}}
	\caption{Keypoint resampling comparisons. (a) An example of distorted point clouds, (b-d) the corresponding keypoints resampled from (a) by using geometry, color, and normal vectors, respectively.}
	\centering
	\label{resample}
\end{figure*}

\begin{table}[t]
\begin{center}
\captionsetup{justification=centering}
\caption{\textsc{Performance Results of Normal Number on SJTU database.}}
\label{table7}
\scalebox{1.2}{
\begin{tabular}{|c|c|c|c|}
\hline
Numbers & SROCC & KROCC & PLCC \\ \hline
6 & 0.79 & 0.61 & 0.85 \\ \hline
11 & 0.87 & 0.71 & 0.91 \\ \hline
12 & \textbf{0.89} & \textbf{0.72} & \textbf{0.93} \\ \hline
13 & 0.87 & 0.70 & 0.91 \\ \hline
15 & 0.85 & 0.68 & 0.90 \\ \hline
18 & 0.83 & 0.65 & 0.89 \\ \hline
\end{tabular}}
\end{center}
\end{table}

\subsection{Performance of Individual Component}
Since the proposed SGR is composed of three groups of quality-aware features, including geometry density, color naturalness, and angular consistency. It is interesting to validate the performance of each individual feature component. In TABLE \ref{table5}, we conduct the ablation study and show the results. It can be seen that the two components from regional pre-processing are combined to boost the performance. Moreover, further integrating angular consistency leads to the best results of the proposed SGR framework.

Additionally, the keypoint resampling is based on the unique normal vectors, we test the cases by using other information, such as the color attributes and geometry information. The performance results are listed in TABLE \ref{table6}, where the proposed method outperforms the others. One possible explanation may be that the normal vectors can represent the structures of point clouds. Here, we also give the comparisons of keypoint resampling in Fig. \ref{resample}. As can be seen in this figure, we take a point cloud distorted by severe Gaussian noise as an example. Compared with the keypoints resampled from geometry or color signals, by using the unique normal vectors, the keypoints are sampled more uniformly and cover a larger range of visual content. Besides, since the extracted quality-aware features after regional pre-processing are based on geometry and color aspects, our method could achieve the disentanglement of the known information to some extent. All these validate the superiority of our proposed structure guided resampling. Furthermore, the used normal number is explored in our experiment. TABLE \ref{table7} shows that 12 normal numbers can deliver the best performance. With more and less normal numbers, the results would decrease. Therefore, we choose this parameter equaling to 12 in the proposed method.

\section{Conclusion}
In this paper, we develop a general and efficient blind/no-reference quality assessment method for 3D point clouds, which is based on structure guided resampling. Inspired by the human perception of typical visual distortions, the proposed SGR model operates for distorted 3D geometry and associated attributes information without any special access to the reference, where regional pre-processing, quality-related feature extraction, and quality regression are involved in the proposed framework. In the experiments, we demonstrate that our proposed SGR algorithm can correlate well with human quality ratings on serveral subject-rated point cloud quality databases. Further, we also show the effectiveness of each constituted component of our proposed method.

In the future, we plan to explore more powerful distortion-aware features to improve the quality assessment model. Moreover, the way to apply our proposed method to the automatic optimization of existing 3D point cloud processing algorithms could be another direction.

\bibliographystyle{IEEEtran}
\bibliography{references}

% biography section
%
% If you have an EPS/PDF photo (graphicx package needed) extra braces are
% needed around the contents of the optional argument to biography to prevent
% the LaTeX parser from getting confused when it sees the complicated
% \includegraphics command within an optional argument. (You could create
% your own custom macro containing the \includegraphics command to make things
% simpler here.)
%\begin{IEEEbiography}[{\includegraphics[width=1in,height=1.25in,clip,keepaspectratio]{mshell}}]{Michael Shell}
% or if you just want to reserve a space for a photo:

%\begin{IEEEbiography}{Michael Shell}
%Biography text here.
%\end{IEEEbiography}
%
%% if you will not have a photo at all:
%\begin{IEEEbiographynophoto}{John Doe}
%Biography text here.
%\end{IEEEbiographynophoto}
%
%% insert where needed to balance the two columns on the last page with
%% biographies
%%\newpage
%
%\begin{IEEEbiographynophoto}{Jane Doe}
%Biography text here.
%\end{IEEEbiographynophoto}

% You can push biographies down or up by placing
% a \vfill before or after them. The appropriate
% use of \vfill depends on what kind of text is
% on the last page and whether or not the columns
% are being equalized.

%\vfill

% Can be used to pull up biographies so that the bottom of the last one
% is flush with the other column.
%\enlargethispage{-5in}

% that's all folks
\end{document}